\documentclass[aps,preprint,epsf]{revtex4} 
\def \bfgr #1{ \mbox {{\boldmath $#1$}}}
\newcommand{\be}{\begin{eqnarray}}

 \usepackage{graphicx}

\newcommand{\ee}{\end{eqnarray}}
\newcommand{\la}{\langle\,}
\newcommand{\ra}{\,\rangle}
\newcommand{\qmod}{|{\bf q}|}
\newcommand{\bfP}{{\bf P}}

\def\btau{{\mbox{\boldmath$\tau$}}}

\begin{document}
 
 \title{Final state interaction effects in semi-exclusive DIS  off the deuteron}
 \author{C. Ciofi degli Atti}
 \author{L.P. Kaptari}
 \altaffiliation{On leave from  Bogoliubov Lab. Theor. Phys.,141980, JINR,  Dubna, Russia}
  \affiliation{Department of Physics, University  of Perugia and INFN Sezione di Perugia,
       via A. Pascoli, Perugia, I-06123, Italy}
\author{B.Z.  Kopeliovich}\altaffiliation[Also at ]{Fleorov Lab. Nucl. Problems. 141980, JINR, Dubna, Russia}

\affiliation{Max Planck Institut f\"ur Kernphysik Postfach 103980, 69029  Heidelberg,  FRG
\\ Institut f\"ur Theoretische Physik der Universit\"at,  93040  Regensburg, FRG}

\vskip 3mm
\date{\today}
\vskip 3mm

 \begin{abstract}
The effects of  the final 
state interaction (FSI) in semi-exclusive
 deep-inelastic scattering of electrons 
   off the  deuteron are analyzed paying particular  attention to  two extreme 
  kinematical regions: i) the one   where  FSI effects are minimized, so that the 
   quark distribution of  bound nucleons could be investigated, and 
  ii) the one where  the  re-interaction  of the produced hadrons with the
  spectator nucleon is maximized, which would allow one 
  to study the mechanism of  hadronization
   of highly virtual quarks.
    It is demonstrated   that when the recoiling spectator nucleon is detected 
     in the backward hemisphere
   with low  momentum,  the effects from the FSI are negligible, 
    whereas at large transverse momenta of the spectator,
  FSI effects are rather large.  Numerical estimates  show that the 
  FSI corrections are sensitive to the theoretical  models of   the
  hadronization mechanism.        
 \end{abstract}
 \maketitle
 \section{introduction} 
 According to  QCD,  deep-inelastic scattering (DIS) is the process in which 
  an incident electron interacts with a  target quark by exchanging
  a gauge boson, making the quark  highly virtual.
   The formation of the final, detectable  hadrons, occurs after the  
   space-time propagation of the created nucleon debris,  
  with a sequence of soft and hard production  
  processes. The theoretical description of these  processes, which generally  cannot 
  be treated within
  perturbative QCD,  requires the use  of model approaches.
  Most of them are   based  upon  the quark color string model \cite{string}, 
  according to which
at  world interval of the order of $\simeq 1 fm$, \,  the string  which is  formed by the   highly  virtual
  leading quark  and the remnant  target quarks,  breaks into  a hadron and another,
   less stretchy
  string. Further,  at longer space-time intervals,  this decay process iterates 
   unless the energy  of the string  is too low for   hadron 
   production and                       
  all the final hadrons are formed.  However, since the  hadronization process can
  also  
  be accompanied  by  gluon 
  perturbative bremsstrahlung \cite{gluon}, the string model itself is not sufficient for
  a consistent treatment of hadronization. A reliable 
  model must incorporate both the perturbative and the non-perturbative aspects of the 
   hadron formation process. 
   Note, that the hadronization process starts at extremely short space-time intervals, 
  hence a direct experimental study  of these intervals is difficult to undertake in DIS off a free nucleon.
  As a matter of fact,    
  the final hadrons do not carry much information
  about their early stage of hadronization, and therefore  only nuclear targets, which consist
  of a number of scattering centers, allows one  to probe short times after
   the hadronization has started.
 In a nucleus, at each hadronization point one  expects 
  re-interactions of the produced hadrons  with the nuclear constituents, 
  so that  the multiplicity of final particles is predicted to be reduced relative
  to the case of nucleon targets.
  Thus, by comparing the same DIS process off a single nucleon
  and off nuclear targets, information on the space-time structure of the hadronization 
  process could be obtained.
    The  theoretical model of  hadronization developed in Ref.\cite{kope1},
  proved to be very effective for the explanation  of the  leading hadron  multiplicity ratios (nucleus
  to nucleon)  measured at HERMES
   \cite{HERMES}  in semi-inclusive processes.
  It should however  be pointed out that   the initial stage of hadronization
  is difficult to investigate  by semi-inclusive  processes, where  the non leading
  hadrons are strongly affected by subsequent cascade processes and therefore do not carry 
  information  on  their  formation mechanism.
   Recently, it has been shown \cite{bocla}
   that
  the deep inelastic semi-exclusive process $A(e,e',(A-1))X$, where the nucleus $(A-1)$ is detected  
  in coincidence with the scattered electron, originally proposed to investigate 
    the  structure function of bound nucleons\cite{tagging}, could  be  an effective tool to study
    the mechanisms of  hadronization and the  initial stage of hadronization.
The simplest of the processes $A(e,e', (A-1))X$, namely that with a deuteron target, {\it viz}
 \be
e+ D = e'+ X + N
\label{reaction}
\ee 
 has been the object
of many theoretical calculations, mainly aimed at studying the neutron structure function
\cite{simula}-\cite{alex}, whereas its experimental investigation is planned to be performed at JLab \cite{kuhn}.
  
  Process (\ref{reaction}) has many attractive features with respect to the inclusive process
  $^2H(e,e')X$. As a matter of fact,  it should be stressed
 that in spite of the fact that
  inclusive DIS processes   have  provided us in the past  with 
  fairly precise knowledge of parton distributions in hadrons, conclusive
  information about the origin of the EMC effect is still lacking; moreover, important details
  on the neutron structure function are unknown, which  is 
  mostly due  the difficulties and ambiguities  related to the  unfolding of the neutron
  structure functions from  nuclear data \cite{ambig}. Semi-exclusive 
  processes could provide, on the contrary, unique information on both the origin
  of the EMC effect, and  the details of the neutron structure function;    
  moreover, they  can also be used as  a  unique tool to investigate   
  hadronization processes. Obviously, a reliable treatment of semi-exclusive processes
  requires a careful treatment of  the  FSI of the nucleon  debris $X$ with the final nuclear system $(A-1)$.
  Intuitively, one expects, on one hand,  that   if the proton is detected
in the backward hemisphere, FSI effects should not play a relevant role, so that
the process could be used to investigate the bound nucleon  structure functions;  on  the other hand,
the effects from FSI are expected to be relevant in the process when the
 recoiling nucleon is detected  in the direction perpendicular to the three-momentum
transfer, in which case information on the  hadronization mechanism could be obtained.
In both cases,  a quantitative estimate
of FSI is a prerequisite for obtaining a reliable  estimate of either nucleon structure functions  or
hadronization  processes. 
In view of the planned experiments at JLab, a quantitative calculation of FSI effects in process
(\ref{reaction}) is called for.  
It is precisely the aim  of the present 
paper to illustrate the results of the calculations of  process (\ref{reaction})
in various kinematical conditions taking FSI into account.

Our paper is organized as follows. In Section \ref{sec:dva}  the  kinematics
and the general formula for the cross section are presented; the theoretical reaction mechanism 
 for the considered process is illustrated  in Section \ref{sec:tri}  both in the  Plane Wave Impulse 
 Approximation (PWIA) (\ref{subsect:triA}) and  taking into account  FSI effects  (\ref{subsect:triB}).
  In the same Section (\ref{subsect:triC})
 the effective cross section describing  the rescattering   of the nucleon debris with the 
spectator nucleons is illustrated. 
Eventually, the results of the numerical calculations are presented and discussed
 in Section  \ref{sec:res}.

\section{kinematics and cross section}\label{sec:dva}
The general theoretical formalism of  process (\ref{reaction}) can be found in
several papers (see e.g. \cite{simula},\,\,\cite{wally},\, and  \cite{tagging}),
therefore  only few general aspects of the problem  will be recalled here.

 In one-photon-exchange approximation, the   cross section for the process (\ref{reaction}) can be written as follows
\be &&
\frac{d\sigma}{dx dQ^2\ d^3p_s}=\frac{4\alpha^2}{Q^4}\frac{\pi\nu}{x}
\left[ 1-y-\frac{Q^2}{4E^2} 
\right]
\widetilde{l}^{\mu\nu}L_{\mu\nu}^D\equiv
\label{eq1}\\&& 
\frac{4\alpha^2}{Q^4}\frac{\pi\nu}{x}
\left[ 1-y-\frac{Q^2}{4E^2} 
\right]
\left [
\tilde l_L W_L +\tilde l_{T} W_T+\tilde l_{TL} W_{LT}\cos\phi_s+
\tilde l_{TT} W_{TT}\cos (2\phi_s) 
\right ]
\label{cross}
\ee
\noindent where  $\alpha$ is the fine-structure constant,
 $Q^2 =-q^2= -(k-k')^2 = {\bf q}^{\,\,2} - \nu^2=4 E E^{'}
sin^2 {\theta \over 2}$ the four-momentum transfer (with ${\bf q} =
{\bf k} - {\bf  k}'$, $\nu= { E} - {E}^{'} $ and $ \theta \equiv
\theta_{\widehat{\vec k \vec k^{'}}}$), $ x = Q^2/2M\nu $ the
Bjorken scaling variable,  
 $y=\nu /E$,   
 ${\bf p}_s$ the momentum of the detected recoiling  nucleon (also called the  {\it spectator} ($s$)
 nucleon), $\widetilde {l}_{\mu\nu} $ and $L_{\mu\nu}^D$ the electron 
 and deuteron electromagnetic tensors, respectively;  the former has the well known standard form, whereas
 the latter can be written as follows 
\be
&&
 L_{\mu\nu}^D=
  \sum\limits_{X}\la {\bf P}_D|J_\mu|{\bf P}_f\ra\la {\bf P}_f|J_\nu|{\bf P}_D\ra 
 \delta^{(4)}\left(k+P_D-k'-p_X-p_s \right)
 d\btau_X,
\label{eq5} \ee where $J_\mu$ is the operator of the deuteron 
electromagnetic current, and ${\bf P}_D$ and ${\bf P}_f= {\bf 
p}_X+{\bf p}_s$ denote the three-momentum of the initial deuteron 
and the final hadron system, respectively, with ${\bf p}_X$ being the 
momentum of the undetected hadronic state created by the DIS process on the active nucleon. In Eq. 
(\ref{cross}) the various $W_i$ represent the nuclear response 
functions and 
\be
\tilde l_L=\frac{Q^2}{\qmod^2},\quad
\tilde l_T=\frac{Q^2}{2\qmod^2} +\tan^2\left(\frac{\theta}{2}\right),
\tilde l_{LT}=\frac{Q^2}{\sqrt{2}
\qmod^2}\sqrt{\tan^2\left(\frac{\theta}{2}\right)+\frac{Q^2}{\qmod^2} },
\tilde l_{TT}=\frac{Q^2}{2\qmod^2}.
\label{eq7}
\ee
 
It is well known, that within the PWIA, i.e., when FSI effects are disregarded,
the four response functions in Eq. (\ref{cross}) can be represented only via 
 two independent structure functions, {\it viz}  $W_L$ and $W_T$.
 Moreover, in the DIS kinematics, when 
  the Callan-Gross  relation holds, i.e. $2x F_1^N(x) =  F_2^N(x)$ ( $F_{1,2}^N(x)$
   are the nucleon DIS
  structure functions in the Bjorken limit),  the semi inclusive 
   cross section (\ref{cross}) will depend 
   only upon one DIS structure function, e.g.
    $F_2^N(x)$.  In the presence of  FSI all four responses contribute to the cross section  (\ref{cross});
  however if  FSI effects  are not too large, nucleon 
 momenta are sufficiently small and the momentum 
 transfer large enough,  one can expect that the 
 additional two structure functions are  
small corrections, so that the  semi-exclusive
 DIS  cross section can  still be described by
  one, effective structure function $F_{2A}^{s.e.}(Q^2,x,p_s)$ 
(\cite{simula,wally,tagging})  
 (the two structure functions $W_{LT}$ and $W_{TT}$
 can exactly be   eliminated in the parallel kinematics , i.e., by  choosing
 the spectator nucleon along the  momentum transfer $\bf q$, i.e.,
$\theta_s=0$ or  $\theta_s=\pi$, or  integrating  the cross section  over $\phi_s$
 \footnote{Integration over $\phi_s$ eliminates from Eq. (\ref{cross})
 the $LT$ and $TT$ structure functions. Alternatively, the  
 $LT$ and $TT$ structure functions 
can be eliminated by choosing the reaction plane at, e.g.,  $45^o$ with respect 
to the scattering one
and performing two measurements of $p_s$ with $\Delta\phi_s = 90^o$.}).
 
  The cross section (\ref{cross}) can be also   represented in terms of  
 the light cone variable $\alpha_s$, $p_{\parallel}$ and $p_{T}$, defined as
\be
\alpha_s=\frac{E_s-p_{\parallel}}{m},\quad
p_{\parallel}= |{\bf p}_s|\cos \theta_s ,\quad
    p_{T}= |{\bf p}_s|\sin \theta_s 
\label{eq8}
\ee
\noindent where $\theta_s$ is the angle between ${\bf p} _s$ and ${\bf q}$ 
 and the spectator four-momentum is $p_s\equiv (E_s, {\bf p}_s)$ 
 (note, that in the  DIS kinematics the  light cone $z$-axis is directed opposite to the
vector ${\bf q}$). 
In terms of the above  variables  the  cross section 
will read as follows
 
 \be
 &&
\frac{d\sigma}{dx dQ^2\ d\alpha_s {dp_T}^2}=
\frac{4\alpha^2}{Q^4}\frac{\pi^2m\nu}{x}
\left[ 1-y-\frac{Q^2}{4E^2} 
\right]      
\left [\phantom{\!\!\!\!\!\!\!\!\frac{m^1}{E_s}}
\tilde l_L W_L +\tilde l_{T} W_T\right] .
\label{eq9}
\ee

\section{The spectator mechanism}\label{sec:tri} 
In what follows we will consider the reaction (\ref{reaction}) within the so called
$ spectator\,\, mechanism$, according to which the DIS process occurs on the so called $\it active$
nucleon, e.g. nucleon "1",  and the second nucleon (the  $ spectator$)
 recoils and is detected in coincidence with the scattered electron.
At high values of the  3-momentum transfer $\bf q$
the produced hadron 
 debris propagates mostly  along the $\bf q$ direction and  re-interacts
 with  the spectator nucleon.  The wave function  of such a state
 can be written in the form
\be &&
\Psi_f (\{ \xi \},{\bf r}_x,{\bf r}_s ) =  \ 
\phi_{\beta_f}(\{ \xi\} )\psi_{p_x,p_s}({\bf r}_x,{\bf r}_s),
\label{eq14}
\ee 
where ${\bf r}_x$ and ${\bf r}_s$  are the coordinates of the
center-of-mass of   system $X$ and the spectator nucleon, respectively, 
$\{ \xi\} $ denotes the set of the internal coordinates of  system $X$,
described by the internal wave function 
$\phi_{\beta_f}(\{ \xi\})$,  $\beta_f$ denoting  all  quantum numbers
 of the final state $X$; 
the wave function  $\psi_{p_x,p_s}({\bf r}_x,{\bf r}_s)$  describes  the relative
motion of  system $X$ interacting elastically  with  the spectator $s$.  
The matrix elements in  Eq. (\ref{eq5}) can  easily be  computed, provided
 the  contribution of the two-body  part of the deuteron electro-magnetic current
can be disregarded, which means that  the deuteron 
current  can be  represented as a sum of electromagnetic 
 currents of individual nucleons,
 $J_\mu (Q^2,X)=j_\mu^{N_1} + j_\mu^{N_2} $.
Introducing  complete sets of plane wave states
 $|{\bf p}_1{'},{\bf p}_2{'} \ra$ and $|{\bf p}_1,{\bf p}_1\ra$ in intermediate states,  one obtains

\be&&
\la {\bf P}_D| j_\mu^N|\beta_f,{\bf P}_f={\bf p}_x+{\bf p}_s\ra=\nonumber \\&&
\sum\limits_{\beta,{\bf p}_1{'},{\bf p}_2{'}}
\sum\limits_{{\bf p}_1,{\bf p}_2}
\la {\bf P}_D|{\bf p}_1{'},{\bf p}_2{'}\ra
\la {\bf p}_1{'},{\bf p}_2{'}|j_\mu^{N_1}| \beta, {\bf p}_1,{\bf p}_2\ra
\la\beta, {\bf p}_1,{\bf p}_2|\ \beta_f,{\bf p}_x,{\bf p}_s
\ra =\nonumber\\&&
\int \frac{d^3 p}{(2\pi)^3} \psi_D({\bf p})
\la {\bf p}|j_\mu^{N_1} (Q^2,p\cdot q)| \beta_f, {\bf p}+{\bf q}\ra
\psi_{\bfgr{\kappa}_f}({\bf p}+{\bf q}/2),
\label{eq15}
\ee
 where $\bfgr{\kappa}_f=({\bf p}_x-{\bf p}_s)/2$.
 In Eq. (\ref{eq15}) the matrix element 
 $\la {\bf p}|j_\mu^{N_1}(Q^2,k\cdot q)| \beta_f, {\bf p}+{\bf q}\ra$
 describes the electromagnetic transition from a moving nucleon 
 in the initial  state to a final hadronic system $X$ in a quantum state $\beta_f$. 
 Obviously, the sum over all the final state $\beta_f$ of the
  square of this matrix element, complemented with the corresponding energy conservation 
  $\delta$-function, generates the deep inelastic 
  nucleon hadronic tensor of a moving nucleon.  
 
\subsection{The PWIA}\label{subsect:triA} 
In PWIA, $\gamma^*$ interacts with a quark of the neutron, a nucleon debris is formed and 
the proton recoils without interacting with the debris. 
The relative motion debris-proton is thus 
described by a plane wave 
\be &&
\psi_{\bfgr{\kappa}_f}({\bf p}+{\bf q}/2)\sim
(2\pi)^3\delta^{(3)} ({\bf p}+{\bf q}/2-\bfgr{\kappa}_f)=
(2\pi)^3\delta^{(3)} ({\bf p}-{\bf p}_s)
\label{eq16}
\ee
and the well known result
\be
&&
\frac{d\sigma}{dx dQ^2\ d\alpha_s dp_T^2}
=
K(x,Q^2,p_s)\, n_D(|{\bf p}|) \, F_2^{N/A}(Q^2,x,p_s),
\label{crossPWIA}
\ee
is obtained, where ${\bf p} = -{\bf p}_s$ is the momentum of the struck nucleon before interaction, 
$K(x,Q^2,p_s)$  a kinematical factor (see, e.g. ref. \cite{tagging}),
$F_2^{N/A}(Q^2,x,p_s) = 2x F_1^{N/A}(Q^2,x,p_s)$  the DIS structure
function of  the  active nucleon,
and $n_D$  the momentum distribution of the hit nucleon, i. e.
\be
n_D(|{\bf p}|)=\frac13\frac{1}{(2\pi)^3} \sum\limits_{{\cal
M}_D} \left |\int d^3 r 
 \Psi_{{1,\cal
M}_D}( {\bf r})\exp(-i{\bf p r}/2) \right|^2.
\label{dismom}
\ee
In general,  since the active nucleon  inside the deuteron is off-mass shell,
 its deep inelastic electromagnetic tensor  has a more complicate structure 
 \cite{kaptoff,thomas} containing off mass shell  corrections. 
However, at low values of the 3-momentum of the spectator, i.e., at low 3-momenta
of the active nucleon before interaction,
these corrections are negligible  and one can safely express   the nucleon tensor
 only in terms of the  DIS structure function  $F_2^{N/A}(Q^2,x,p_s)$.

\subsection{FSI}\label{subsect:triB} 
  Consider now  FSI effects  within  the   kinematics 
when  the   spectator is slow,  
 and  the momentum transfer large enough so that 
 the rescattering process of the fast system $X$ off the spectator
  $s$ could be considered as a high-energy soft hadronic interaction.   
 In this case the momentum of the detected spectator ${\bf p}_s$ 
 only slightly differs from the  momentum ${\bf p}$ before rescattering, so that
  in  
 the matrix element  
  $\la {\bf p}|j_\mu^{N_1} (Q^2,p\cdot q)|\beta_f, {\bf p}+{\bf q}\ra$
   one can take  ${\bf p}\sim -{\bf p}_s$,  obtaining
in co-ordinate space 
\be
&&
\la \bfP_D| j_\mu^N|\bfP_f\ra\cong j_{\mu}^N(Q^2,x, {\bf p}_s)
\int d^3 r 
 \psi_D({\bf r}) \psi_{\bfgr{\kappa}_f}^+({\bf r})\exp(i{\bf r q}/2).
\label{eq23}
\ee
 The cross section then becomes

\be
&&
\frac{d\sigma}{dx dQ^2\ d\alpha_s dp_T^2}
=
K(x,Q^2,p_s)\, n_D^{FSI}({\bf p}_s,{\bf q}) \, F_2^{N/A}(Q^2,x,p_s),
\label{crossfsi}
\ee
where  
\be
n_D^{FSI}({\bf p}_s,{\bf q})=\frac13\frac{1}{(2\pi)^3} \sum\limits_{{\cal
M}_D} \left |\int d^3 r 
 \Psi_{{1,\cal
M}_D}( {\bf r}) \psi_{\bfgr{\kappa}_f}^+({\bf r})\exp(i{\bf r q}/2) \right|^2,
\label{dismomfsi1}
\ee
\noindent is the distorted momentum distribution, which coincides with the 
 momentum distribution of the hit nucleon
(Eq.(\ref{dismom}))
 when $\psi_{\bfgr{\kappa}_f}^+({\bf r})\sim \exp(-i\bfgr{\kappa}_f\bf r)$ .

 All the effects from the FSI in Eq. (\ref{crossfsi}) are 
implicitly generated by  the
dependence of the distorted momentum distribution (\ref{dismomfsi1}) upon ${\bf q}$, and
by the features of the rescattering  wave function $\psi_{\bfgr{\kappa}_f}^+({\bf r})$;
the latter 
describes  the  rather complicated 
many-body  system  $X + spectator$ and can only be generated by model approaches.
 However, in our particular case, when the relative momentum  $\bfgr \kappa_f \sim  {\bf q}$
 is rather large and the rescattering processes occur with low momentum transfers, the
 wave function $\psi_{\bfgr{\kappa}_f}^+({\bf r})$
  can be replaced by its eikonal form describing 
  the propagation of the 
 nucleon  debris formed after  $\gamma^*$ absorption by a target quark, followed  by  hadronization processes and interactions
 of the newly produced  pre-hadrons with the spectator nucleon. This series of soft
 interactions with the spectator can be characterized by an effective cross
 section  $\sigma_{eff}(z,Q^2,x)$  depending upon time (or the distance $z$ traveled  by the system $X$).
Thus the distorted  nucleon momentum distribution  (Eq. \ref{dismomfsi1}) becomes

\begin{equation}
 n_D^{FSI}( {\bf p}_s,{\bf q}) =
\frac13\frac{1}{(2\pi)^3} \sum\limits_{{\cal
M}_D} \left | \int\, d  {\bf r} \Psi_{{1,\cal
M}_D}( {\bf r}) S( {\bf r},{\bf q}) \chi_f^+\,\exp (-i
{\bf p}_s {\bf r}) \right |^2,
 \label{dismomfsi}
\end{equation}
where $\chi_f$ is the  spin function of the spectator nucleon and 
$S( {\bf r},{\bf q})$   the $S$-matrix describing 
the  final state interaction
between the debris and the  spectator,  {\it viz.}  
\begin{equation}
S({\bf r},{\bf q}) = 1-\theta(z)\, \frac{\sigma_{eff}(z,Q^2,x)(1-i\alpha)}{4\pi b_0^2}\,
\exp(-b^2/2b_0^2).
 \label{gama}
\end{equation}

\noindent where the $z$ axis is directed along ${\bf q}$, i.e. 
${\bf r} = z \displaystyle\frac{\bf q}{|{\bf q}|}+\bf b$. 

As previously mentioned, such a treatment  of FSI holds if,
during the process of hadronization,  the
FSI occurs as a series of soft, elastic scattering of the
produced debris
 with the spectator nucleon and the kinematics is
 chosen in such a way, that 
the DIS  structure function of the nucleon  
does not change too rapidly. 
In this case  the factorization assumption  holds and   
the rescattering process is described as the debris-nucleon rescattering via an  effective 
 cross section. Hence, the usual
eikonal approximation can be applied. 
Note, that the above expressions  can also  be used to calculate the effects of FSI 
in quasi-elastic  scattering (QES)
by simply replacing the debris-nucleon cross section 
with the z-independent nucleon-nucleon cross section (see, e.g., \cite{nikolaev,ckt}).

\subsection{The effective  debris-nucleon cross section}
\label{subsect:triC} 
 Recently in Ref.\cite{bocla}
 the final state interaction  in  DIS off nuclei
 due to the propagation of the struck nucleon debris and its hadronization in
  the nuclear environment, has been
 considered  and applied to the process  of the type  $A(e,e'(A-1))X$ with $A \geq 4$.
  This process precisely coincides with the one investigated  in the present paper, 
 i.e the process $^2H(e,e'p)X$.  The basic ingredient 
 governing  the FSI, i.e.  the time- and $Q^2$-dependent effective cross section $\sigma_{eff}$
describing the interaction of the debris with a nucleon of the spectator system
 $(A-1)$, has been
obtained in Ref. \cite{bocla}
 on the basis of a model  which takes into
account both the production of hadrons due to the breaking of the color
string, which is formed after a quark is knocked out off a bound nucleon,
as well as the production of hadrons originating from gluon radiation.
The general expression has the following form

 \be
\sigma_{eff}(t)=\sigma^{NN}_{tot} +
\sigma^{\pi N}_{tot}\Bigl[n_M(t) +
n_G(t)\Bigr]\ ,
\label{bb.6}
 \ee
  where $\sigma^{NN}_{tot}$ and $\sigma^{\pi N}_{tot}$ 
  are the total cross sections of nucleon-nucleon and 
  meson-nucleon interaction, and  $n_M(t)$ and $n_G(t)$ are the effective
  numbers of created mesons and radiated gluons, respectively, and 
  are explicitly given in Ref. \cite{bocla}. There the color-dipole
   picture was employed by replacing each radiated gluon by a color-octet $q\bar q$
   pair.

 The cross section (\ref{bb.6}) exhibits a rather complex 
 $Q^2$- and $x$-dependence, which, however,   asymptotically  
 tends to a  simple logarithmical behavior.  This is illustrated  
 in Fig.~\ref{Fig1}, where the $Q^2$-dependence  of the effective cross section, calculated with the
 set of parameters given in Ref. \cite{bocla}, is exhibited.

\section{Results and discussion}\label{sec:res}
 Eq.(\ref{crossfsi}) is the basic equation that one has to  evaluate in order to provide a quantitative 
 significant description of the quasi-exclusive process  $A(e,e'(A-1))X$; the latter, as already mentioned, 
 has been originally suggested 
 \cite{tagging} as a powerful tool   to investigate the reaction mechanism of  DIS off nuclei and the 
bound nucleon structure function; it was however soon realized \cite{bocla} that it  could also  be
an extremely powerful tool to investigate the hadronization mechanism.
In the present paper we will focus on   both aspects of the problem, namely 
 we will address the problem of finding  two different kinematics, 
 a one in which  FSI corrections  are negligibly  small, 
 so that a direct  study of the DIS
 nucleon (neutron)  structure functions becomes possible, and another one, where
 FSI are maximized, thus obtaining information  on the hadronization mechanism.
 As it can be seen  from  Eq. (\ref{crossfsi}),  all of the FSI effects are contained  in the
 distorted momentum distribution  $ n_D^{FSI}( {\bf p}_s, {\bf q})$ (Eq. (\ref{dismomfsi})),
  so that the investigation of its   deviation from the 
 nucleon momentum distributions  ($n_D (|{\bf p}|)$)) could 
   provide clear 
   signature on  FSI effects. This is   illustrated in Fig. \ref{Fig2}, which shows
 the ratio
$n_D^{FSI}/n_D$, (with  $n_D^{FSI}$ and $n_D$  given
by Eqs. (\ref{dismomfsi}) and (\ref{dismom}), respectively)  calculated   using two  different 
models for the effective cross section $\sigma_{eff}$, representing its  
upper and lower limits, {\it viz} the time- and $Q^2$-
dependent cross section  of Ref. \cite{bocla} (solid lines), and 
a constant cross section $\sigma_{eff}=20\,\, mb$ 
(dashed lines) considered in Ref. \cite{wally}. 
In our  numerical  calculations the parameters entering  Eq. (\ref{gama}), {\it viz}
the slope  $b_0$  and the ratio $\alpha$ of the real to the
imaginary part of the forward amplitude, have been  taken from $NN$ 
scattering data
at high energies.
It should be pointed out, in this respect, that our results, in the range of considered momenta, 
are not very sensitive to the value of $\alpha$; whereas the values of the latter 
is known in the case of {\it nucleon-nucleon} scattering, the value for  
{\it debris-nucleon} scattering  should rely on some theoretical models. This
point is under investigations and the results will be presented elsewhere.         
 It can be seen from 
Fig.~\ref{Fig2}
 that the predictions given by the two different 
models of the effective cross section ere rather  different,
 particularly when the recoiling proton is emitted perpendicularly to ${\bf q}$, i.e. 
 at $\theta_s \sim 90^o$,  and with large values of the 
momentum ($p_s\sim 0.2\,\, GeV/c$). 
 Thus, by investigating this  kinematical  region  one could, in principle,  obtain 
unique information about the magnitude of  $\sigma_{eff}$
and, consequently, about the hadronization mechanism.

In order to illustrate the different role played by the FSI in Deep Inelastic and Quasi-Elastic
scattering, we show in  Fig. \ref{Fig3}  the ratio $n_D^{FSI}/n_D$
for the QE process $^2H(e,e'n)p$, which corresponds to the calculation
of $n_D^{FSI}$ using  the effective nucleon-nucleon cross section, i.e.
$\sigma_{eff}=44\,\, mb$. It can be seen, in agreement with the results for heavier nuclei \cite {bocla},
 that the survival probability of $(A-1)$ in DIS is less than in QE scattering.
     
The results exhibited in Figs.  \ref{Fig2} and \ref{Fig3} also demonstrate that FSI effects are essentially reduced
in  parallel kinematics ($\theta=0^o, 180^o$)  and also at small values of $p_s$. 
This region is attractive from the   experimental point of view since 
  in such a kinematics 
the   structure functions $W_{LT}$ and $W_{TT}$ in
Eq. (\ref{cross}) vanish exactly and the effective structure function $F_2$ in
Eq.  (\ref{crossPWIA}) exactly coincides with the DIS structure function of a bound nucleon.  
Thus, in this kinematical region a direct experimental study of the off-mass  
shell effects in DIS becomes possible. The results exhibited in Figs. \ref{Fig2} and \ref{Fig3}
 have been
obtained by using the deuteron wave function corresponding to  the Reid Soft Core (RSC)
 potential \cite{RSC}.
At small values of $|{\bf p}|$ different  potential models 
provide basically the same  deuteron wave function, which is not true, however, 
  at 
moderate and large values of $|{\bf p}|$.
Therefore, we have investigated the   dependence of the ratio  $n_D^{FSI}/n_D$
upon the two body potential used to generate the deuteron wave function.
The results are presented  in
Figs. \ref{Fig4}  where the predictions of the Reid and Bonn \cite{bonn} potentials are
compared in correspondence of two  different values of the transverse momentum $p_T$ (cf.Eq. \ref{eq8}).
It can be  seen that, independently of  the emission angle of the spectator nucleon, 
the two potentials provide essentially   the same results at relatively small 
  values of the spectator momenta $|{\bf p}_s|$, but very   different results at
  large  values of $|{\bf p}_s|$  ($|{\bf p}_s|\ge 0.25 GeV/c$). 

\begin{figure}[t] 
  \vskip 5cm
\begin{center}
    \includegraphics[height=0.35\textheight]{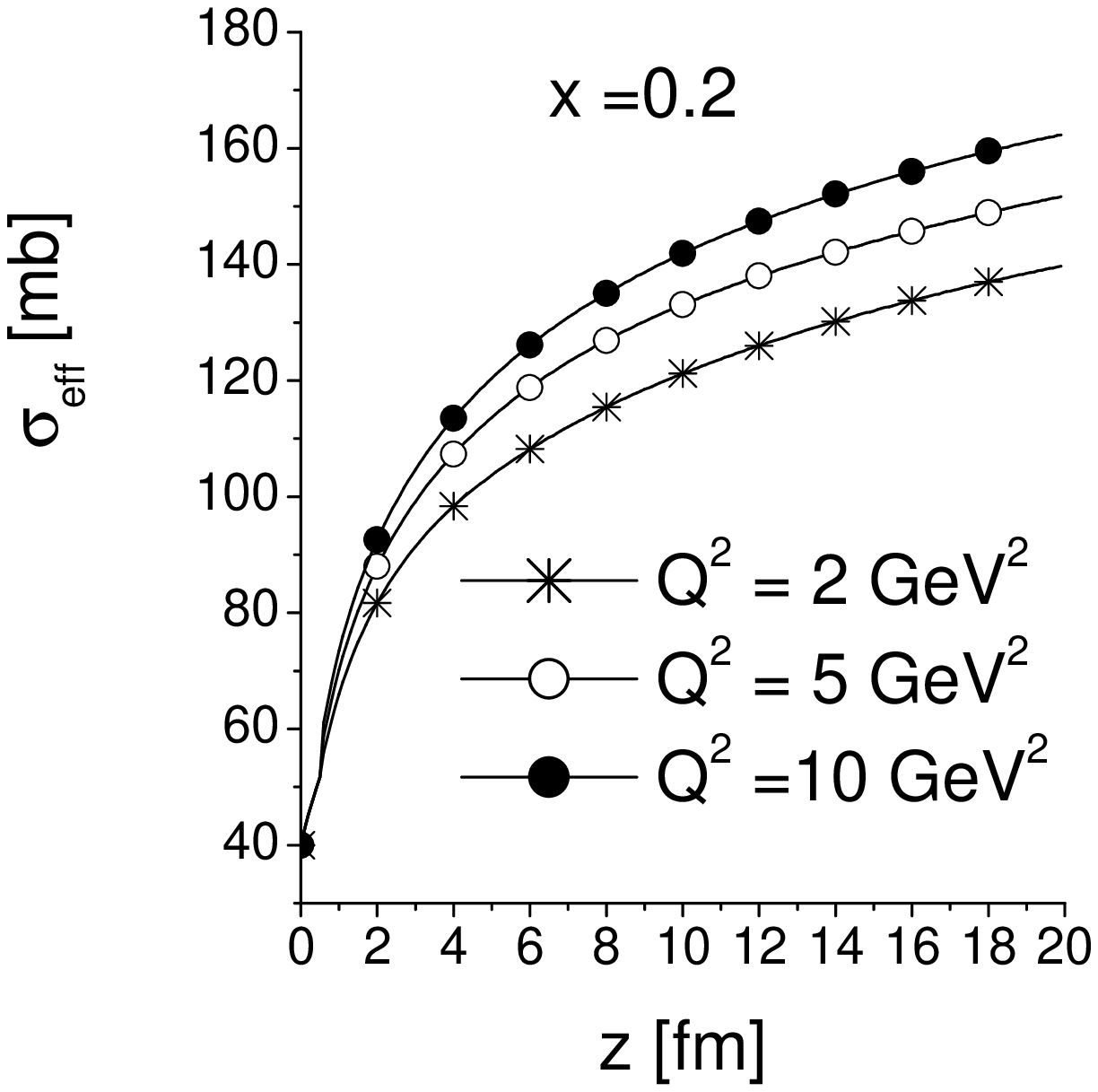}
\caption{The debris-nucleon effective cross section (Eq. \ref{bb.6}) plotted {\it vs} the distance $z$ for a fixed 
value of the Bjorken scaling variable  $x$ and various values of the four-momentum transfer 
 $Q^2$ (after  Ref. \cite{bocla})}
    \label{Fig1}

\end{center}
\end{figure}

\begin{figure}[h] 
   \vskip 5cm                        
\includegraphics[height=0.8\textheight]{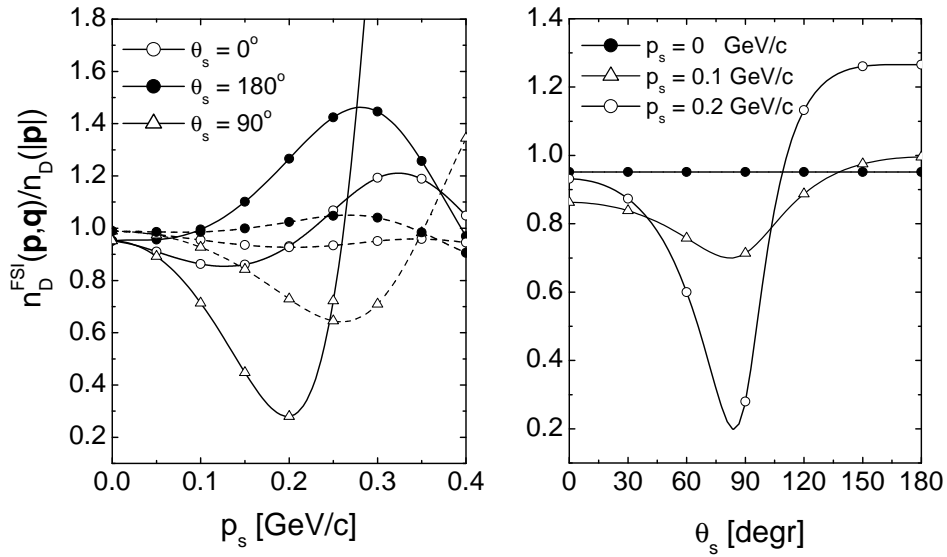} 
\vskip -10cm
\caption{ {\it Left panel}: the Deep Inelastic Scattering ratio  $n_D^{FSI}/n_D$
with $n_D^{FSI}$ and $n_D$ given, respectively, by Eqs. (\ref{dismomfsi}) and (\ref{dismom}),
 calculated
{\it vs.} the momentum  $p_s \equiv |{\bf p}_s|$ of the spectator nucleon emitted at different angles $\theta_s$.
The full lines correspond to the  $Q^2$- and $z$-dependent   debris-nucleon
effective cross section $\sigma_{eff}$ shown in Fig. \ref{Fig1}, whereas the
 dashed lines correspond to a constant cross section $\sigma_{eff} = 20\,\, mb$. {\it Right panel}: the same as in 
 the left panel, {\it vs} the spectator emission angle ${\theta_s}$ for different values of the
 spectator momentum. Calculations have been performed at  
 $Q^2=5\,\, (GeV/c)^2$ and $x=0.2$.}
\label{Fig2}  
\end{figure}

\begin{figure}[ht] 
   \vskip 5cm  
\includegraphics[height=0.8\textheight]{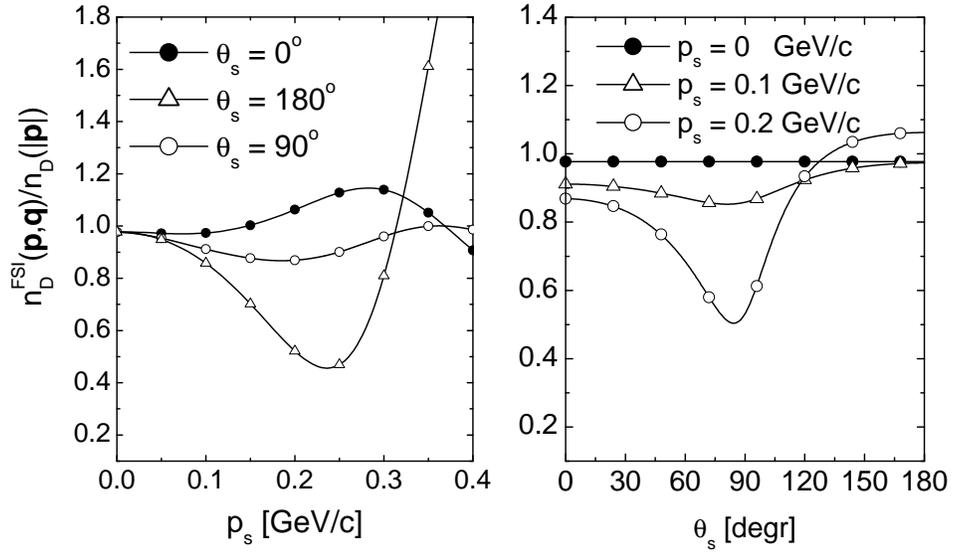} 
 \vskip -10cm
\caption{  {\it Left panel}: the Quasi-Elastic  Scattering ratio  $n_D^{FSI}/n_D$
with $n_D^{FSI}$ and $n_D$ given, respectively, by Eqs. (\ref{dismomfsi}) and (\ref{dismom}),
 calculated with the nucleon-nucleon   $\sigma_{eff}$, i.e.   $\sigma_{eff} = 44 \,\,m b$,
{\it vs.} the momentum  $p_s \equiv |{\bf p}_s|$ of the spectator nucleon emitted at different angles $\theta_s$.
  {\it Right panel}: the same as in 
 the left panel, {\it vs} the spectator emission angle ${\theta_s}$ for different values of the
 spectator momentum. Calculations have been performed at  
 $Q^2=5\,\, (GeV/c)^2$ and $x=0.2$ .}
 \label{Fig3}  
\end{figure}

\begin{figure}[t] 
  \vskip 5cm
\includegraphics[height=0.8\textheight]{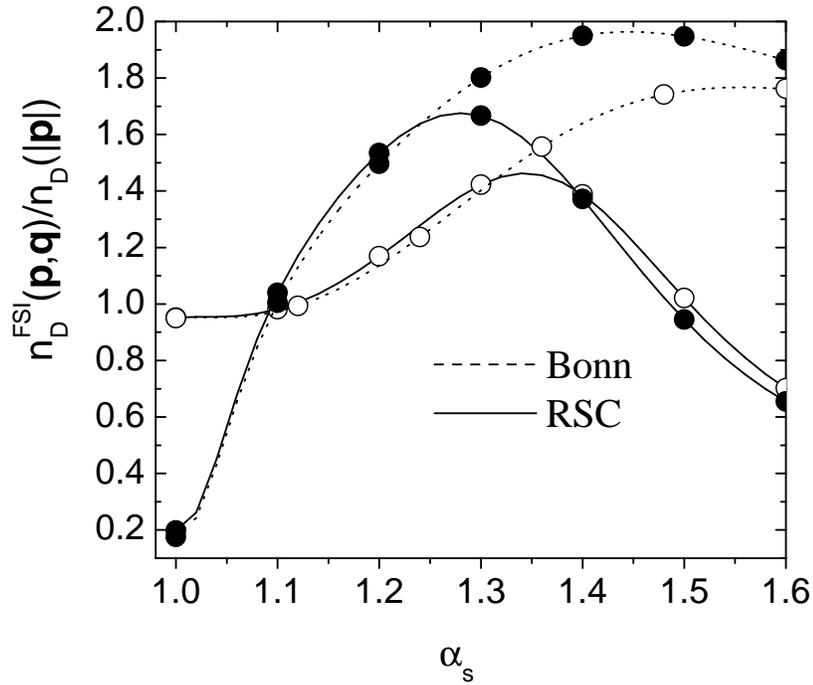} 
\vskip -8cm
\caption{The Deep Inelastic Scattering  ratio $n_D^{FSI}/n_D$ {\it vs} the light cone variable 
 $\alpha_s$ (cf. Eq.\ref{eq8}),
 calculated at $Q^2=5\,\, GeV^2/c^2$
and $x=0.2$ using two different deuteron wave functions: the solid lines correspond to 
the  RSC potential and the dashed lines  to the  Bonn  
potential. Calculations have performed in correspondence of two values of the transverse
momentum $p_T$ of the recoiling nucleon  (cf. Eq. \ref{eq8}), namely  $p_T= 0 \,  GeV/c$ (open dots) and
$p_T= 0.2\  GeV/c$ (full dots). The results correspond to the debris-nucleon cross section  shown in Fig.1. }
\label{Fig4}  
\end{figure}

\begin{figure}[t] 
  \vskip 5cm
\includegraphics[height=0.8\textheight]{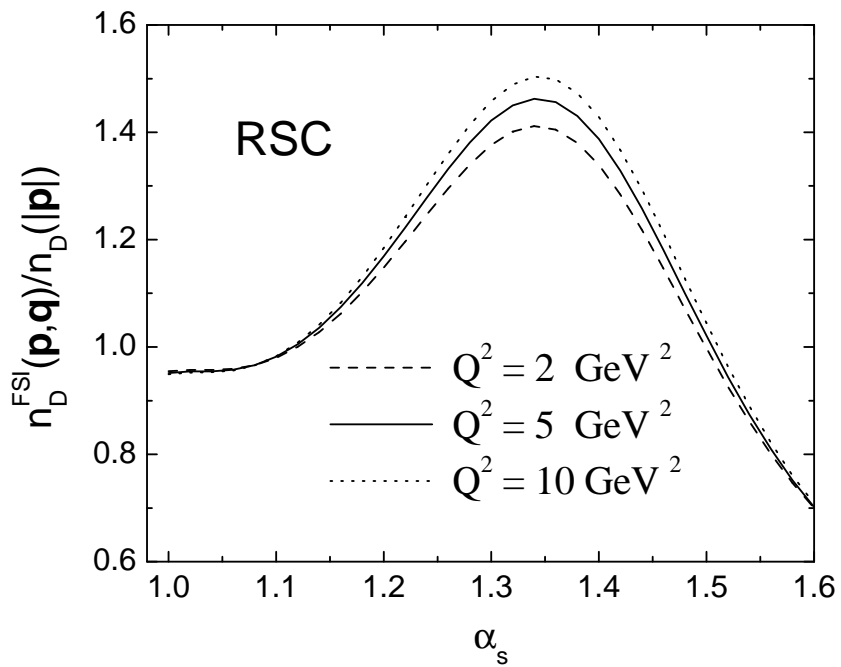} 
\vskip -8cm
\caption{The Deep Inelastic Scattering ratio $n_D^{FSI}/n_D$  {\it vs} the light cone variable
 $\alpha_s$ (cf. Eq.\ref{eq8}),  calculated at different values of $Q^2$ 
and  fixed values  of  $x=0.2$ \, and  $p_T=0\,\ GeV/c$.  The deuteron wave function corresponds to the RSC interaction.}
\label{Fig5}  
\end{figure}

\begin{figure}[t] 
   \vskip 5cm
\includegraphics[height=0.8\textheight]{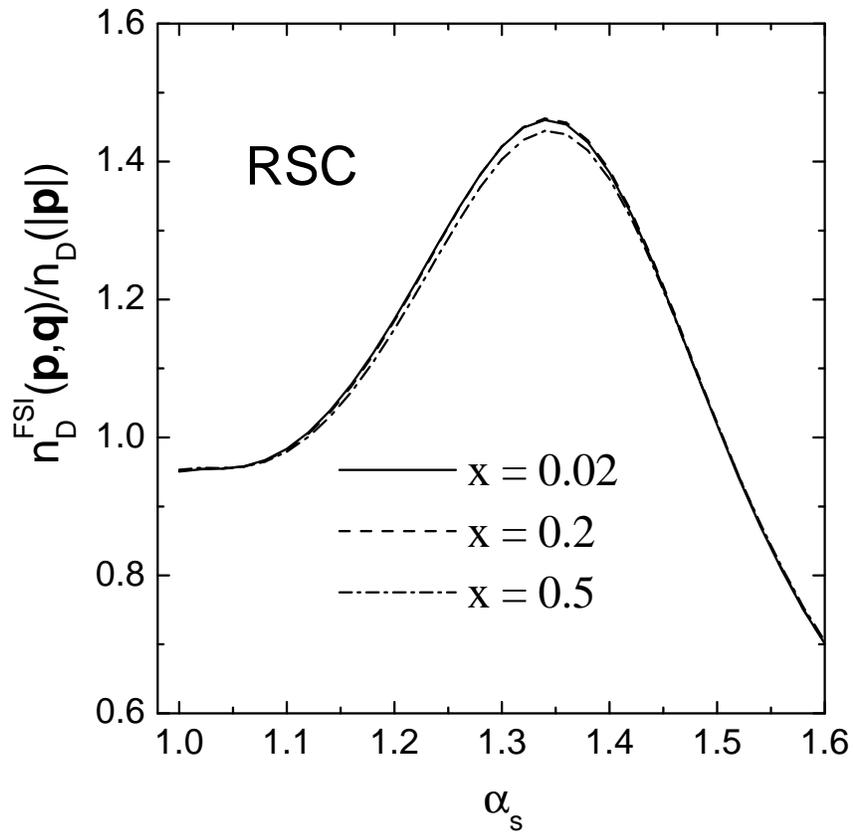}
\vskip -7cm 
\vskip 10mm
\caption{The same as in Fig. \ref{Fig5} for various values of the Bjorken scaling variable $x$ and fixed values of 
 $Q^2=5\,\, (GeV/c)^2$ and $p_T=0 \,\,GeV/c$. The solid and dashed lines practically coincide.}
\label{Fig6}  
\end{figure}

 The dependence of the ratio $n_D^{FSI}/n_D$
 upon the electron kinematics ($Q^2$ and $x$)
 is entirely contained in the $Q^2$ and $x$-dependence of the effective cross section 
 $\sigma_{eff}$  given by  Eq. (\ref{bb.6}). Since this dependence is weak (see Fig. \ref{Fig1}),
 FSI effects in the quasi-exclusive  process $^2H(e,e'p)X$ are 
  practically  $Q^2$ and $x$-independent (see, e.g. Figs. \ref{Fig5} and
 \ref{Fig6}).
 
  To sum up, from the analysis we have exhibited, it can be  concluded  that FSI effects in
   the semi-exclusive process
  (\ref{reaction}) are negligible  in the backward kinematics with slow
   momenta of the detected nucleon, which would allow one to investigate  the nucleon 
   structure function of bound nucleons, in particular the neutron one;  if,  on the contrary,
   the spectator nucleon is emitted perpendicularly to the momentum transfer, FSI effects
   are enhanced and different models for the hadronization process could be investigated.
   
   This work was partially supported by the Ministero dell'Istruzione, Universit\`{a} e Ricerca (MIUR), 
 through the funds COFIN01.
L.P.K. is indebted to  the University of
Perugia and INFN, Sezione di Perugia, for warm hospitality and financial support.


 \newpage  

  
\newpage
 
\newpage

\newpage

\end{document}